\documentclass[graybox]{svmult}

\usepackage{mathptmx}       
\usepackage{helvet}         
\usepackage{courier}        
\usepackage{type1cm}        
\usepackage{makeidx}         
\usepackage{graphicx}        
\usepackage[bottom]{footmisc}

\begin{document}

\pagenumbering{arabic}

\title*{Structure and Evolution of Dwarf Galaxies}
\author{John Kormendy}

\institute{\vskip -5pt Department of Astronomy, University of Texas at Austin, 2515 Speedway, Stop C1400, \phantom{~~~~~~~~} 
           \quad Austin, TX 78712-1205, USA; email: kormendy@astro.as.utexas.edu; ~~~~~~~~~~~~~~~
           Max-Planck-Institut f\"ur Extraterrestrische Physik, Gie\ss enbachstra\ss e, 
           85748 Garching bei \phantom{~~~~~~~~} \quad M\"unchen, Germany; ~~~~~~~~~~~~~~~~~~~~~~~~~~~~~~~~~~~~~~~~~~~~~~~~~~~~~~~~~~~~~~~~~~~~~~~~~~~~~~~~~~~~~~~~~~~
           ~~~~~~~~~~~~~~~~~~~~~ Universit\"ats-Sternwarte M\"unchen, Ludwig-Maximilians-Universit\"at,
           Scheinerstra\ss e 1, \phantom{~~~~~.~~~} \quad\kern-2pt 81679 M\"unchen, Germany}
%
%
\maketitle


\abstract{
Different structural parameter correlations show how classical bulge components and elliptical galaxies are
different from spiral and S0 galaxy disks, irregular (Im) galaxies, and spheroidal (Sph) galaxies.  In contrast, the latter,
apparently diverse galaxies or galaxy components have almost identical parameter correlations.  This shows that they are
related.  A review of galaxy transformation processes suggests that S0 and spheroidal galaxies are star-formation-quenched,
``red and dead'' versions of spiral and Im galaxies.  In particular, Sph galaxies are bulgeless S0s.
This motivates a parallel sequence galaxy classification in which an S0a--S0b--S0c--Sph sequence of decreasing bulge-to-total
ratios is juxtaposed to an \hbox{Sa--Sb--Sc--Im} sequence of star-forming galaxies.  All parameter sequences show a 
complete continuity from giant galaxies to the tiniest dwarfs.  Dwarfs are not a new or different class of galaxies.  Rather, they
are the extreme products of transformation processes that get more important as gravitational potential wells get more shallow.
Smaller Sph and S$+$Im galaxies have lower stellar densities because they retain fewer baryons.  Comparison of
the baryonic parameter correlations with those for dark matter halos allows us to estimate baryon loss as
a function of galaxy mass.  Extreme dwarfs are almost completely dominated by dark matter.
}

\def\etal{et al.~}
\def\gapprox{$_>\atop{^\sim}$} 

\pagestyle{plain}

\section{A Parallel Sequence Galaxy Classification}
\label{sec:1}

      This paper combines visible-galaxy scaling relations from Kormendy \& Bender (2012) with dark matter 
scaling relations from Kormendy \& Freeman (2014).  Results on visible galaxies are conveniently encoded in a
parallel sequence galaxy classification scheme shown in Fig.~1.

\vfill

\includegraphics{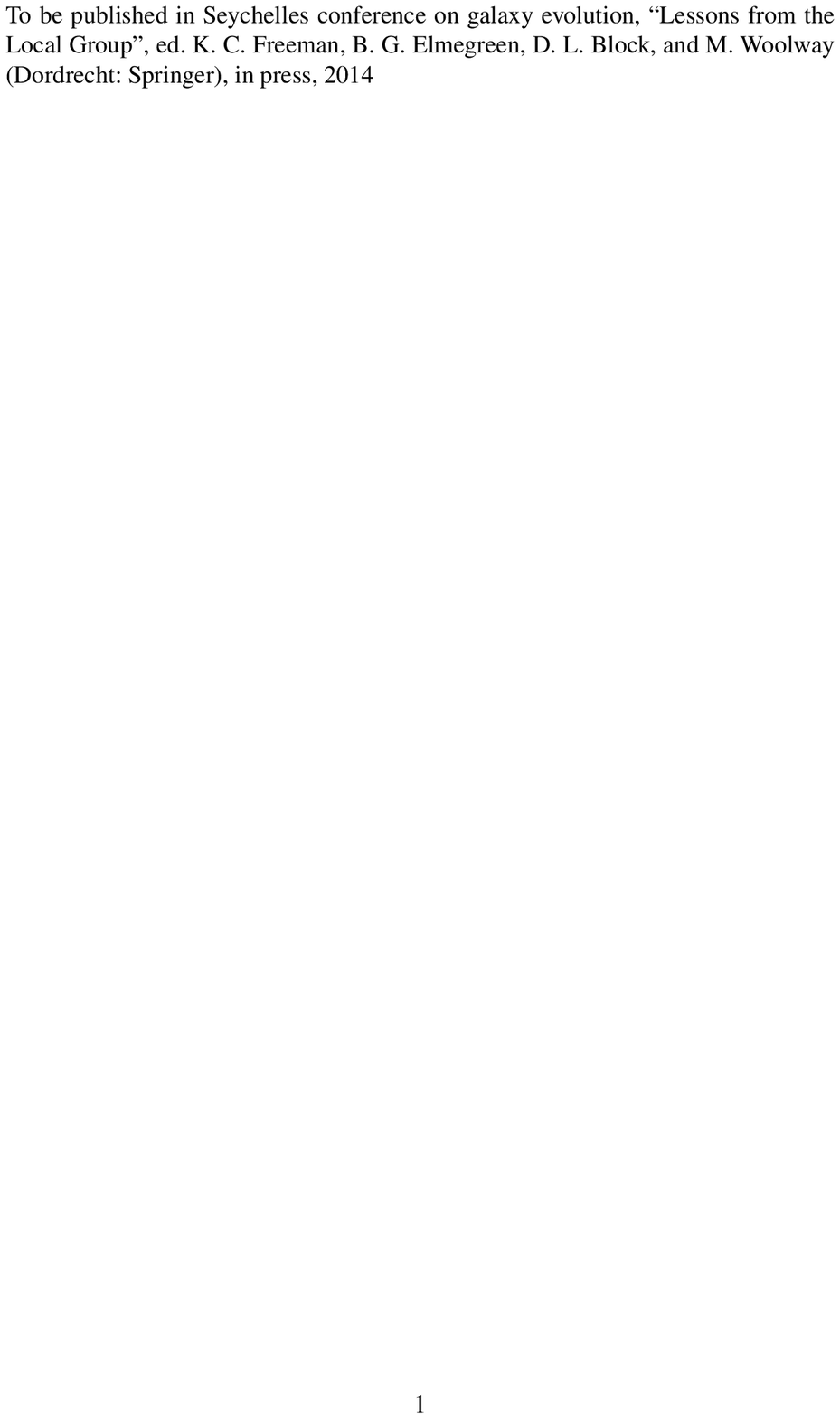}

\includegraphics{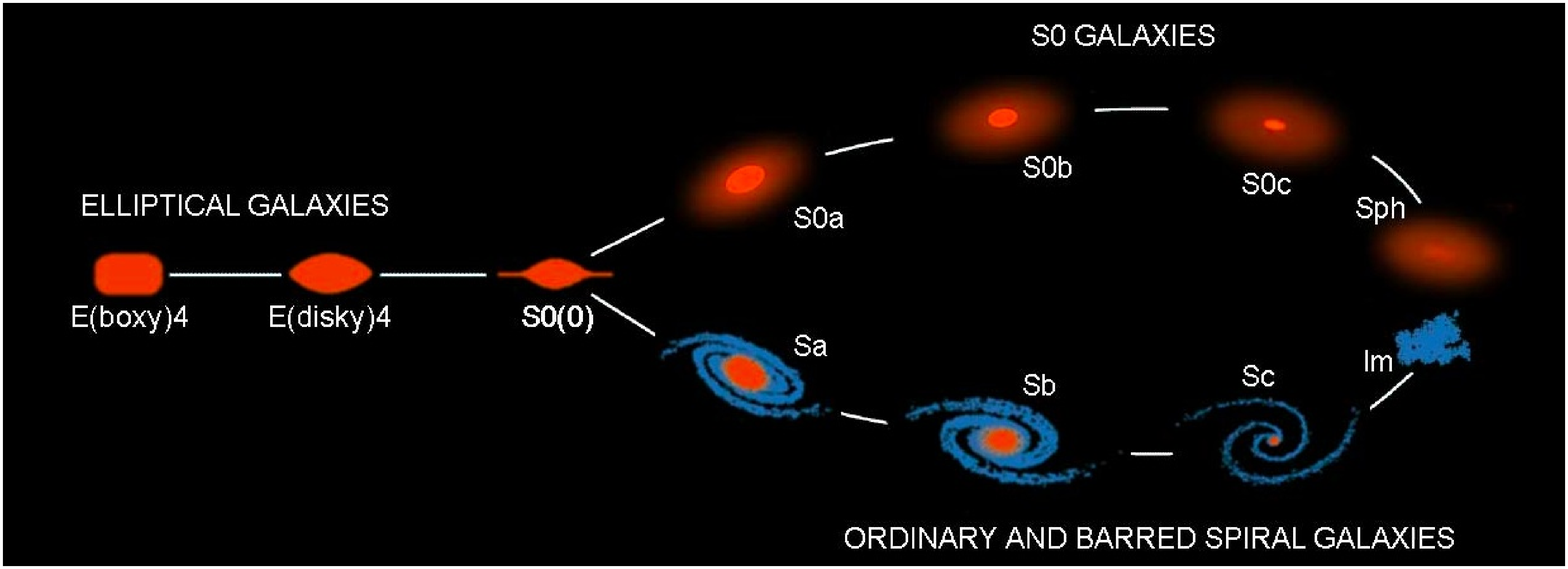}

\noindent {\bf Fig.~1}~~Parallel sequence galaxy classification including spheroidal (Sph) galaxies~as~bulgeless S0
             galaxies juxtaposed with irregular (Im) galaxies.  From Kormendy~\&~Bender~(2012).

\eject

\setcounter{footnote}{0}

\noindent This updates a parallel sequence classification proposed by van den Bergh (1976).  Kormendy \& Bender add
Sph galaxies as S0s that have no bulge component.  This is motivated by the observation that S0 galaxies have bulge-to-total 
luminosity ratios that range from almost 1 to almost 0 (see also Laurikainen \etal 2010).  Kormendy \& Bender find 
a continuous transition in properties -- including density profiles but also kinematic properties -- between 
S0 disks and spheroidal galaxies.  Figure 2 shows the parameter correlations that illustrate this continuity.
S0-like galaxies are given a diffferent name when they have no bulges: They are called Sphs.

Cappellari \etal (2011) and Krajnovi\'c \etal(2011) propose a similar parallel sequence classification (without Sphs). 

\pagenumbering{arabic}
\setcounter{page}{2}

\vfill

\includegraphics{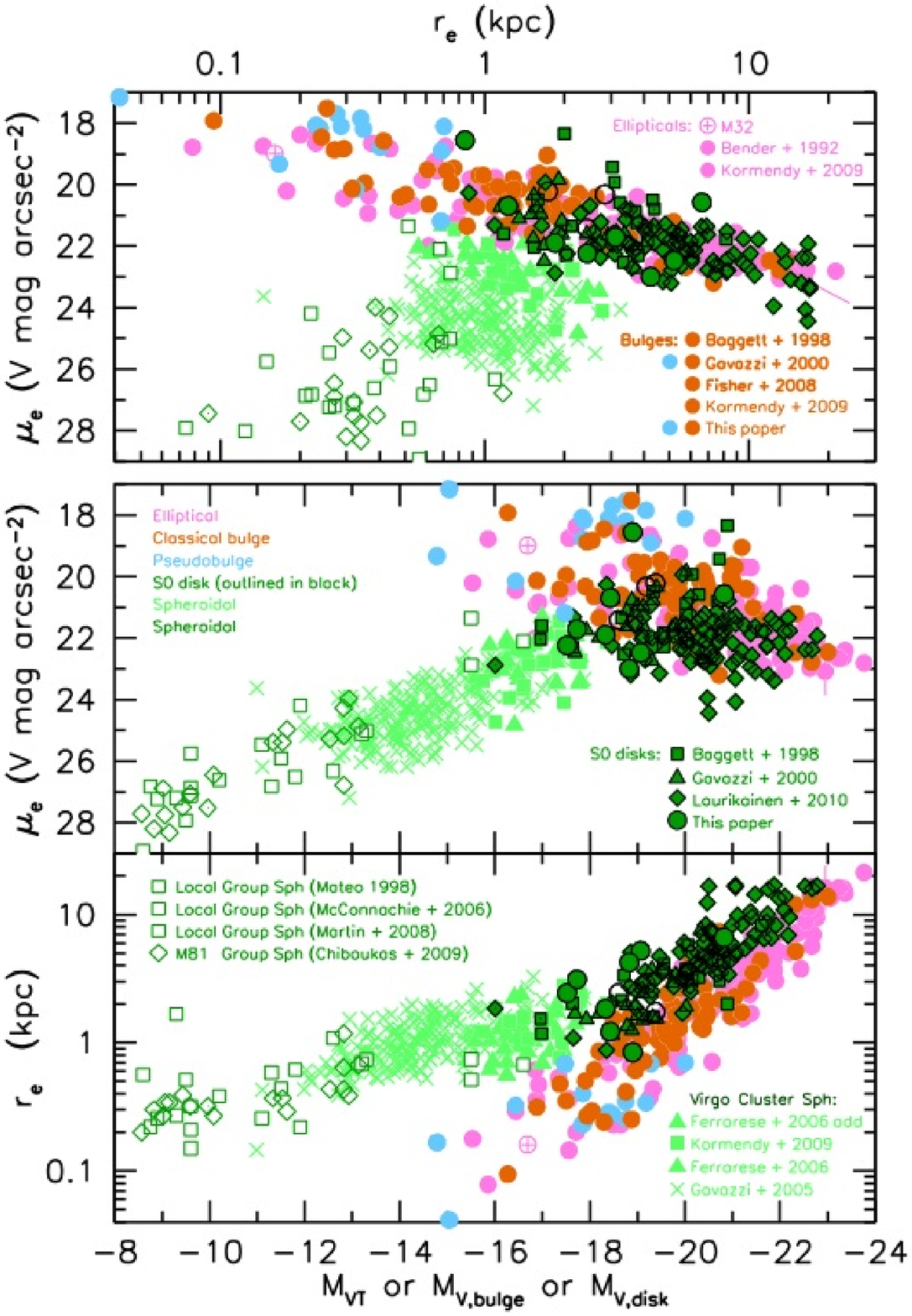}

\noindent {\bf Fig.~2}~~Parameter correlations for ellipticals, bulges, Sph galaxies, and S0 disks.  Bulges and disks 
of S0s are plotted separately.  Plotted parameters are the major-axis effective radius $r_e$ that encloses half of the 
light of the galaxy component, the effective brightness $\mu_e$ at $r_e$, and the total $V$-band absolute magnitude of 
the galaxy or galaxy component.  The middle panel shows the Freeman (1970) result that disks of giant galaxies 
tend to have the same central surface brightness $\mu_0$.  Here, $\mu_e \simeq 22.0$ $V$ mag arcsec$^{-2}$ \hbox{corresponds 
(for an exponential)} to $\mu_0 = \mu_e - 1.82 \simeq 20.2$ $V$ mag arcsec$^{-2}$ $\simeq$ 21 $B$ mag arcsec$^{-2}$, 
brighter than Freeman's value 21.65 $B$ mag arcsec$^{-2}$ because $\mu_e$ is not corrected to \hbox{face-on} orientation. 
From Kormendy \&{ Bender (2012), who conclude:~(1) Sphs are continuous with the disks of S0 galaxies.
(2)~The kink in the $\mu_e$\ts--\ts$M_V$ relation at $M_V \sim -18$, where bulges disappear~in~Fig.~4, 
    marks the transition to a baryon retention sequence: tinier dwarf galaxies retain fewer baryons (Fig.~6).  
Continuity between Sphs and S0 disks is one reason why Fig.~1 shows spheroidal galaxies as bulgeless S0s.
Note that Sphs overlap in $M_V$ with but are distinct from bulges and elliptical galaxies.  They are not ``dwarf ellipticals;''
they are related to disks (Kormendy 1985, 1987). \hyphenpenalty 10000

\eject

\section{S0$+$Sph Galaxies as Transformed Spiral$+$Irregular Galaxies}
\label{sec:2}

      The observational result that motivates the conclusions in this paper is illustrated in Figure\ts3: Samples now of
hundreds of galaxies show:~{\it The continuous parameter sequence defined by Sphs and by the disks (but not bulges) 
of S0 galaxies is indistinguishable from the parameter sequence defined by Magellanic irregular (Im) galaxies and by the disks 
(but not bulges) of spiral galaxies.}  This result was first found for dwarf galaxies by Kormendy (1985, 1987).  
Kormendy \& Bender (2012) extend it to the highest-luminosity S and S0 disks.  

      {\it The most robust conclusion is that Sph galaxies\ts$+$S0 disks are related to Im galaxies{\ts}$+${\ts}S galaxy
disks.  Both together are fundamentally different from classical bulges and ellipticals.}  This conclusion is
also based on additional results that are not included in structural parameter correlations; one is that S$+$S0 disks are 
flat, whereas bulges$+$ellipticals are essentially three-dimensional.  Understanding the above similarities and differences 
requires interpretation, but that interpretation now seems comfortably secure: 
(1) {\it Bulges$+$elliptical galaxies form via major mergers, whereas disks$+$Im$+$Sph
galaxies form by cold gas accretion.} And more directly from Figure\ts3:~(2) {\it Sphs$+$S0 disks are
transformed -- i.{\ts}e., star-formation-quenched -- ``red and dead'' versions of Im galaxies{\ts}$+${\ts}S galaxy disks.}
\hyphenpenalty 10000

\vfill

\includegraphics{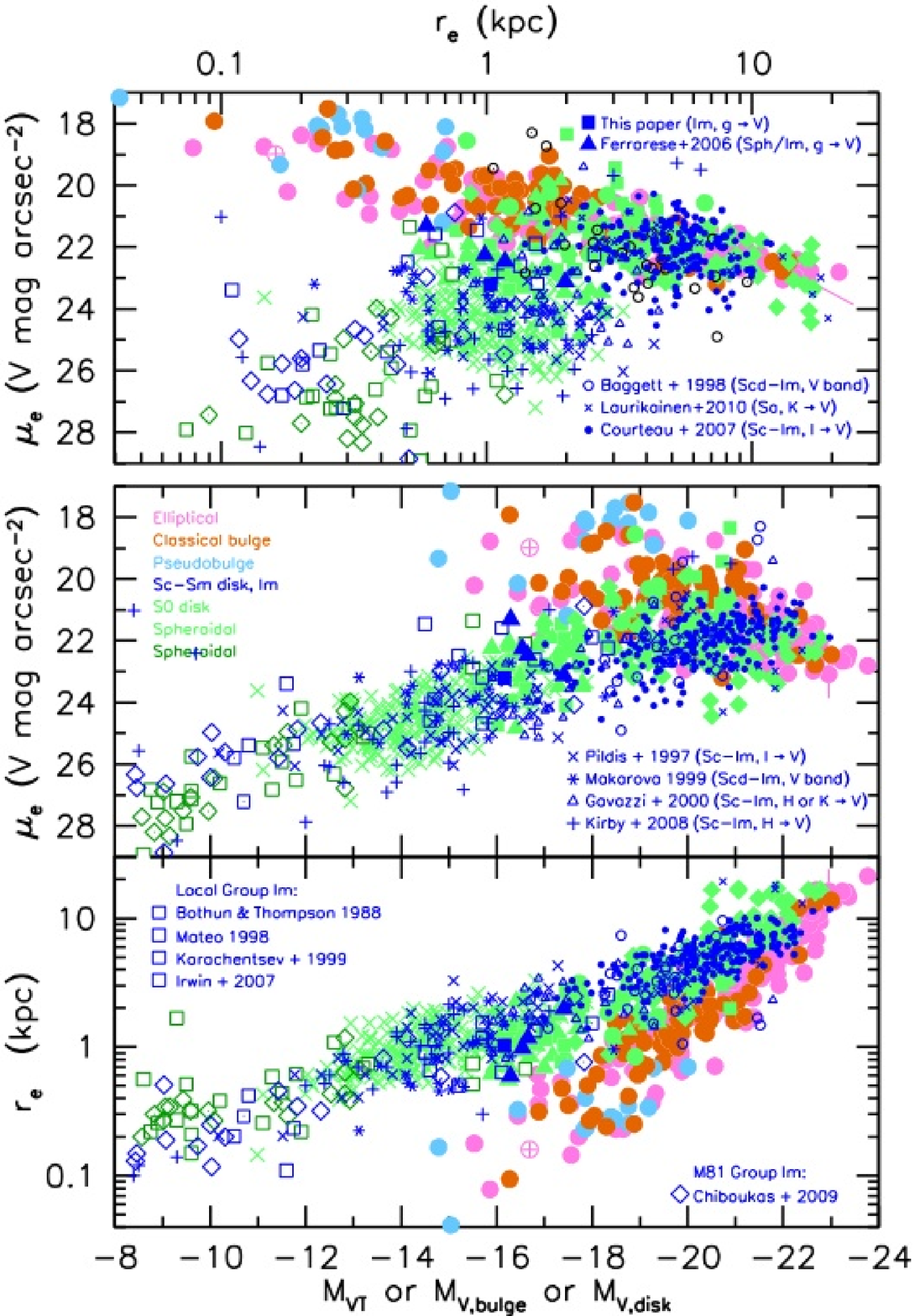}

\noindent {\bf Fig.~3}~~Figure 2 parameter correlations with disks of Sa{\ts}--{\ts}Im galaxies added as blue points.
When bulge-disk decomposition is necessary, the two components are plotted separately.  Disks are not corrected
to face-on orientation.  The blue points represent 407 galaxes from 14 sources listed in the keys.  From Kormendy
\& Bender (2012): They conclude that structural parameter correlations for Sa{\ts}--{\ts}Im galaxy ``disks'' are almost
identical to those for S0\ts--{\ts}Sph ``disks''.  This confirms conclusions in Kormendy (1985, 1987)
using a large galaxy sample.

\eject

\section{S$+$Im $\rightarrow$ S0$+$Sph Galaxy Transformation Processes}
\label{sec:3}

      Kormendy \& Bender (2012) provide an ARA\&A-style review of physical processes that heat galaxy disks
and that use up or remove gas and so quench star formation.  They suggest that the following processes all are
important in transforming Im\ts$+${\ts}S disks into Sph\ts$+${\ts}S0 disks:

\begin{enumerate} \null\vskip -24pt\null
\item{{\it Internal process:}~Dekel \& Silk (1986; see also Larson 1974; Saito 1979) ``suggest that {\it both the 
dIs and the [dSphs] have lost most of their mass} in winds after the first burst of star formation, and that this 
process determined their final structural relations.  The dIs somehow managed to retain a small fraction of their 
original gas, while the [dSphs] either have lost all of their gas at the first burst of star formation or passed 
through a dI stage before they lost the rest of the gas and turned [dSph].''  Kormendy and Bender conclude that 
{\it ``the Sph\ts$+${\ts}Im sequence of decreasing surface brightness with decreasing galaxy luminosity is a sequence 
of decreasing baryon retention''} (emphasis in both originals). \vskip 2pt Figure 2 shows that surface brightnesses in the
Sph\ts$+$S0 sequence start to decrease at $M_{V,\rm disk} \simeq -18$, just where bulges disappear and where we therefore
change galaxy classifications from S0 to Sph (i.{\ts}e., where plot symbols change from dark to light green).
Figure\ts4 independently finds the S0$\rightarrow$Sph transition from rotation curve decompositions.  
As dark matter $V_{\rm circ}$ decreases, bulges decrease in importance relative to disks until they disappear at 
$V_{\rm circ} \simeq 104 \pm 16$ km s$^{-1}$.  The Tully-Fisher (1976) relation shows that this corresponds to
$M_V \simeq -19$ for spirals (Courteau \etal 2007) and $M_V \simeq -18$ for S0s (Bedregal \etal 2006).  It is probably 
not an accident that baryons start to be lost roughly where bulges stop to contribute to the gravitational potential.}
\end{enumerate}

\vfill

\includegraphics{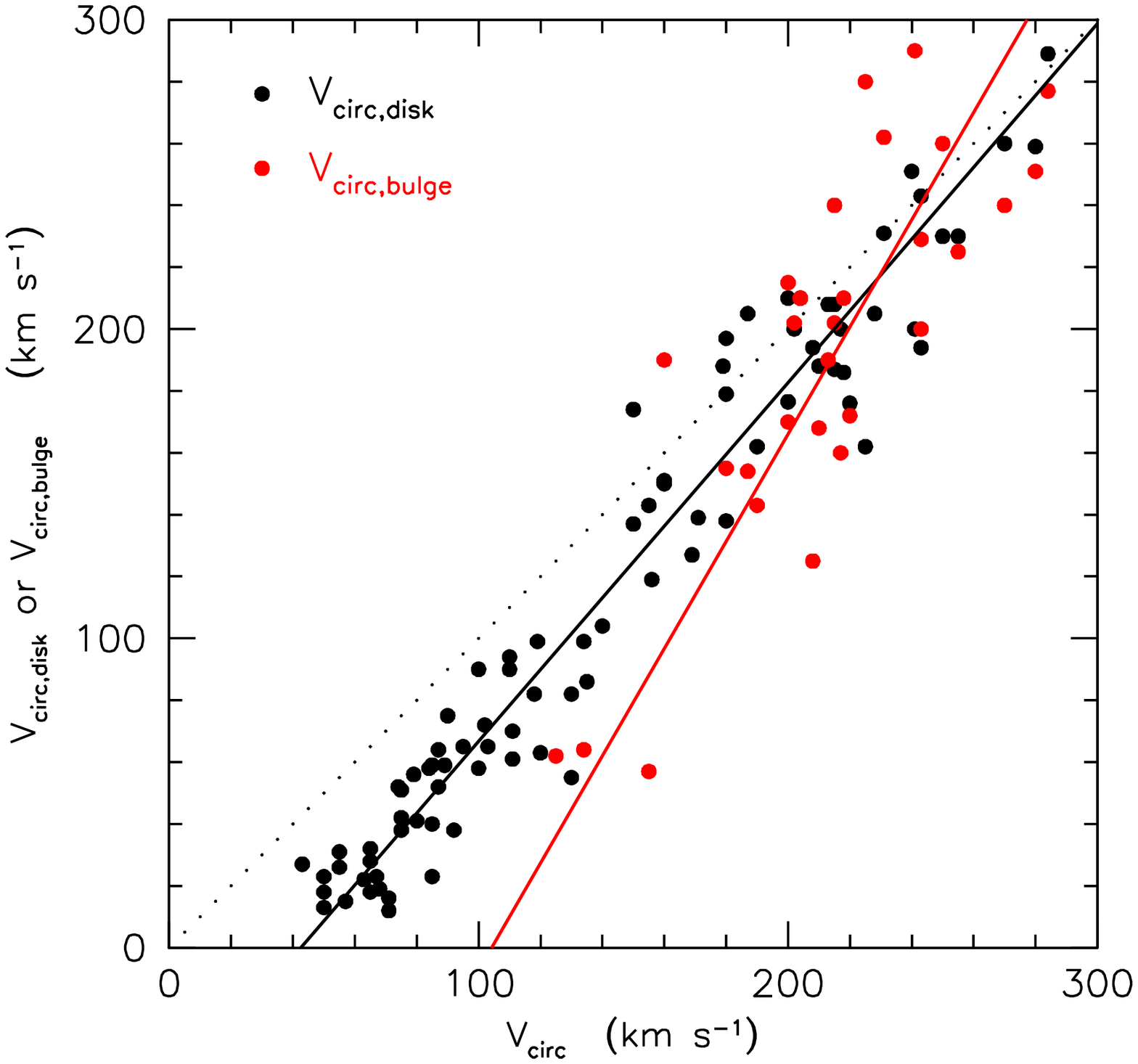}

\noindent {\bf Fig.~4}~~Maximum rotation velocity of bulge ($V_{\rm circ,bulge}$:\ts{\it red points})
and disk $V_{\rm circ,disk}$ ({\it black~points}) components given in bulge-disk-halo decompositions of 
observed rotation curves $V(r)$ whose outer, dark matter rotation velocities~are~$V_{\rm circ}$ 
(Kormendy \& Freeman 2014; references to the $V(r)$ decomposition papers are given there).  The dotted 
line indicates that rotation velocities 
of the visible and dark matter are equal.  Every~red~point has a corresponding black point, but many 
late-type galaxies are bulgeless, and then the plot shows only a black point.  The lines are symmetric
least-squares fits; the disk fit is
$V_{\rm circ,disk}  = (1.16 \pm 0.03)(V_{\rm circ} - 200) + (183 \pm 3)$ km s$^{-1}$;
$V_{\rm circ,bulge} = (1.73 \pm 0.29)(V_{\rm circ} - 200) + (166 \pm 9)$ km s$^{-1}$
is the bulge fit.  The bulge correlation is steeper than the disk correlation. Bulges disappear at 
$V_{\rm circ} \simeq 104 \pm 16$ km s$^{-1}$.  
\lineskip=-8pt \lineskiplimit=-8pt

\eject

\begin{enumerate}
\item{({\it Continued\/})\ts~The transition in Figures 2 and 3 to (I suggest) a baryon retention sequence in smaller dwarfs corresponds
within errors to the transition in edge-on galaxies from giants that do to dwarfs that do not have well-defined dust lanes in their 
disk midplanes (Dalcanton \etal 2004).  They argue that the transition is not caused by changes in gas density.  Rather,
they argue that it is caused by a transition in giant galaxies to a regime in which disk instability leads to lower gas turbulence, 
smaller gas scale heights, and enhanced star formation.~It is plausible that star formation is less efficient~in~smaller~dwarfs. 
However: {\it The observation that dwarf Im and Sph galaxies show the same decrease in surface brightness
with decreasing luminosity shows that the dominant effect is not one that depends on the presence of gas.  This favors the suggestion
that the $\mu_e(M_V)$ relation in Figures 2 and 3 is a baryon retention sequence at $M_V > -18$.}  See also Figure 6.
\vskip 3pt
It may be surprising that Sph$+$S0 and Im$+$S galaxies have similar surface brightnesses, because the latter are star-forming
and so presumably have smaller mass-to-light ratios.  However, (1) at high luminosities, internal absorption partly 
compensates for smaller mass-to-light ratios, and (2) at low luminosities, star formation in Im galaxies is gentle.
Nevertheless, it is fair to emphasize that more work is needed to understand the detailed engineering that results in Sph$+$S0 and Im$+$S
correlations that look so nearly identical.} \vskip 3pt

\item{{\it Environmental processes I:\/} Ram-pressure stripping\ts--{\ts}long underestimated in importance\ts-- is now An Idea Whose 
           Time Has Come.  Gunn \& Gott (1972) suggested, based~on~the detection of X-ray-emitting, hot gas in the Coma 
           cluster (Meekins{\ts}et{\ts}al.{\ts}1971;{\ts}Gursky{\ts}et{\ts}al. 1971) that ``{\it a typical galaxy moving in it will be stripped of its interstellar 
material.}~~ We expect {\it no normal spirals\/} in the central regions of clusters like Coma. The lack of such systems is, of course,
observed'' (emphasis in the original).  Calculations persuaded some people to neglect ram-pressure stripping even while many
observations provided indirect evidence for its importance.~Spirals near the center of the Virgo
cluster are deficient in H{\ts}I gas (Cayette \etal 1990, 1994; Chung \etal 2009).  Also, 
Faber \& Lin (1983);
Lin \& Faber (1983);
Kormendy (1987),
van den Bergh (1994b), and
Kormendy \etal (2009, hereafter KFCB)
suggested that Sph galaxies are ram-pressure-stripped dS$+$Im galaxies, based in part on observations
(Einasto \etal 1974;
van den Bergh 1994a, 1994b; 
Mateo 1998)
that\ts-- with a few (understandable) exceptions -- close dwarf companions of Local Group giant galaxies are almost all spheroidals, that
distant companions are irregulars, and that galaxies with  intermediate (Sph/Im) morphologies live at intermediate distances.  Ever since
van den Bergh (1976), these ideas provided the interpretation of a parallel sequence classification (Figure 1 here) that was constructed
operationally to encode the full range of S0$+$Sph bulge-to-total luminosity ratios $B/T$ from almost 1 to exactly 0.
\vskip 3pt
Spectacular observations of ram-pressure stripping in action now clearly demonstrate~the importance of this process (see
Sun{\ts}et{\ts}al.\ts2010; 
Kormendy \& Bender 2012 for reviews).  
Kenney \etal (2004,\ts2008),
Oosterloo \& van{\ts}Gorkom 2005, and
Chung \etal (2007,\ts2009) 
find that many spiral galaxies near the center of the Virgo cluster show long H$\alpha$ or
H{\ts}I tails interpreted to be cold gas that is being stripped by the ambient X-ray-emitting gas (B\"ohringer \etal 1994).
Kormendy and Bender emphasize~that~these galaxies and the H{\ts}I\ts-{\ts}depleted galaxies are substantially brighter than almost all
Sphs: ``If even the deep gravitational potential wells of still-spiral galaxies suffer H{\ts}I stripping, then the shallow
potential wells of dS\ts$+${\ts}Im galaxies are more likely to be stripped.''  The most impressive recent example of ram-pressure
stripping is the multi-wavelength \hbox{(CO{\ts}gas$+$H{\ts}II$+$X-ray)} tail of the 
galaxy ESO 137-001 in Abell 3627 
(Sun \etal 2006, 2007, 2010;
Woudt \etal 2008;
Sivanandam \etal 2010;
Pavel \etal 2014; Figure 5 here). 
\vskip 3pt
For a recent treatment of ram-pressure stripping in the Galactic halo, see Gatto \etal (2013).
}
\end{enumerate}

\centerline{\null} \vskip 2.28truein 

\includegraphics{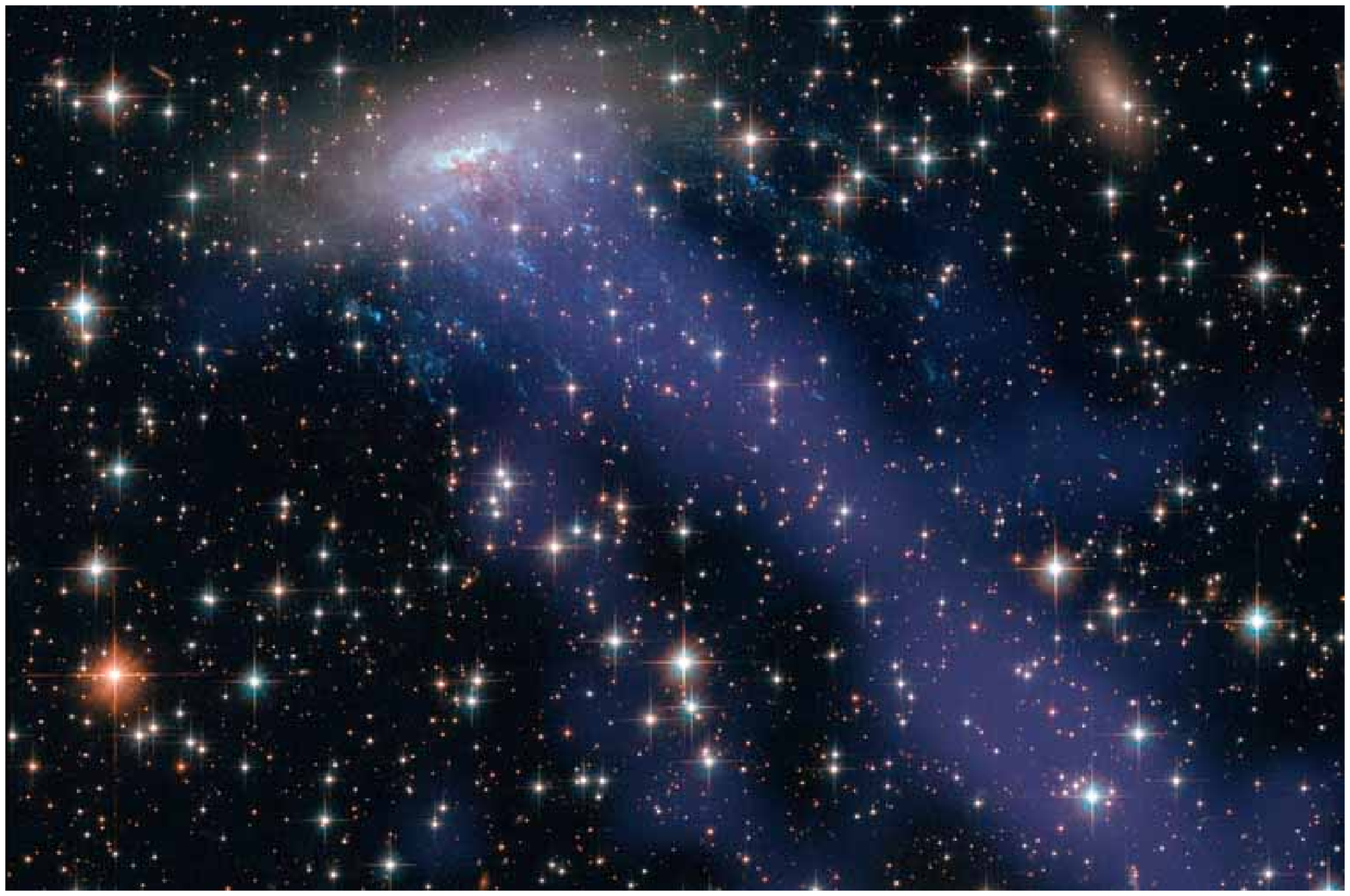}

\noindent {\bf Fig.~5}~~Composite {\it Hubble Space Telescope} (HST) and {\it Chandra X-Ray Observatory\/} image of galaxy
ESO 137-001 in Abell 3627.  The HST image is a $I$-, $g$-, and $U$-band color composite.
Added in blue is the X-ray image; it extends the lighter blue optical streaks of ongoing star formation toward the lower-right.  
This material is interpreted to be ram-pressure stripped. The image source is
{\tt http://www.spacetelescope.org/images/heic1404b/} and {\tt http://apod.nasa.gov/apod/ap140328.html.}
At a distance of $\sim$\ts64 Mpc, the absolute magnitude of ESO 137-001 is $M_V \sim -20.8$.  Conveniently,
this field of view also shows a normal-looking Sph galaxy at upper-right; it is $\sim$\ts3 mag fainter 
than \hbox{ESO\ts137-001.}  Again, if we see a giant galaxy caught in the process of undergoing ram-pressure stripping, 
it~is~not surprising that a much smaller Sph galaxy is thoroughly ``red and dead.''

\vskip 7pt

\item[\llap{3.\kern -2pt}]{{\it Environmental processes II:\/} Dynamical harassment results from many, high-speed encounters with other galaxies 
in a cluster and with the overall cluster potential.  Simulations~show~that (1) it promotes gas flow toward 
galaxy centers, (2) it heats disks, especially vertically, and (3)~it~strips off the outer parts of galaxies
(Moore \etal 1996, 1998, 1999; 
Lake \etal 1998). 
Even in poor environments like the Local Group, tidal stirring of dwarfs on elliptical orbits around the Galaxy or M{\ts31
should have similar effects
(Mayer \etal 2001a,{\ts}b,~2006). 
One success of this picture is that inflowing gas can feed star formation and help to explain why spheroidals,
in which star formation stopped long ago, do not have lower surface brightnesses than current versions of S$+$Im progenitors 
(Figure 3).  This process is clean and inescapable.\vskip 3pt

      Kormendy \& Bender (2012) conclude that {\it dynamical harassment is much more important in the Virgo cluster
than we thought.}  Edge-on S0s with close companions show warps in their outer disks that will phase-wrap 
around the center into structures that resemble Sphs in their shapes and density profiles (NGC\ts4762 and NGC\ts4452; 
their Figures 4 and 8, respectively).  They identify S0/Sph transition objects: 
(1) NGC 4638 is an edge-on S0 with a Gaussian (i.{\ts}e., radially truncated) inner disk embedded in a boxy, E3
    halo that has the properties of a large Sph (their Figures\ts13\ts--\ts16).
(2) VCC\ts2048 is an E6 Sph with an embedded edge-on, S0 disk (their Figures\ts10\ts--\ts12).  
Indeed, many Virgo S0s have Gaussian disk profiles.  Most telling is the observation that several
Virgo cluster S0s have bars embedded in steep-, often Gaussian-profile lenses with no sign of a disk
outside the bar (NGC 4340, NGC 4442, and NGC 4483, their \S{\ts}A.12).  We do not know how to form a bar
that fills the whole disk, because an outer disk is generally required as an angular momentum sink to allow a bar to grow.
Kormendy and Bender suggest that the outer disks in these galaxies have been heated or stripped off.  Finally, they interpret 
the ``new class of dwarfs that are of huge size and very low surface~brightness'' (Sandage \& Binggeli 1984) 
as ``spheroidals that have been harassed almost to death.'' \vskip 4pt

\item[\llap{4.\kern -2pt}]{{\it Environmental processes III:\/} Starvation of late growth by cold-gas infall (Larson \etal 1980)
seems inevitable in environments
like the center of the Virgo cluster where the ambient gas is very hot.  Even in environments like the Local Group, Sph
galaxies that orbit around the Galaxy and M{\ts}31 at velocities  $V \sim 200$ km s$^{-1}$ are unlikely to
encounter cold gas slowly enough to be able to accrete it.  And much of the gas in the Local Group may in any case be
in a warm-hot intergalactic medium (WHIM: Dav\'e \etal 2001). \vskip 4pt

      So Kormendy \& Bender (2012) ``suggest that the relevant question is not `Which of these mechanisms is correct?'
It is `How can you stop any of them from happening?'  It seems likely that all of the above processes matter.''
Engineering details probably depend on environment.

\null \vskip -25pt \null

\section{Visible Matter and Dark Matter Parameter Correlations}
\label{sec:4}

      Figure 6 illustrates our conclusion that, at $M_V$\ts\gapprox\ts$-18$, the dwarf spiral, Im, and Sph
galaxies in earlier figures form a sequence of decreasing baryon retention in smaller galaxies 
(KFCB;
Kormendy \& Bender 2012;
Kormendy \& Freeman 2014).
In contrast, bulges and ellipticals \phantom{000000000}

\vfill

\includegraphics{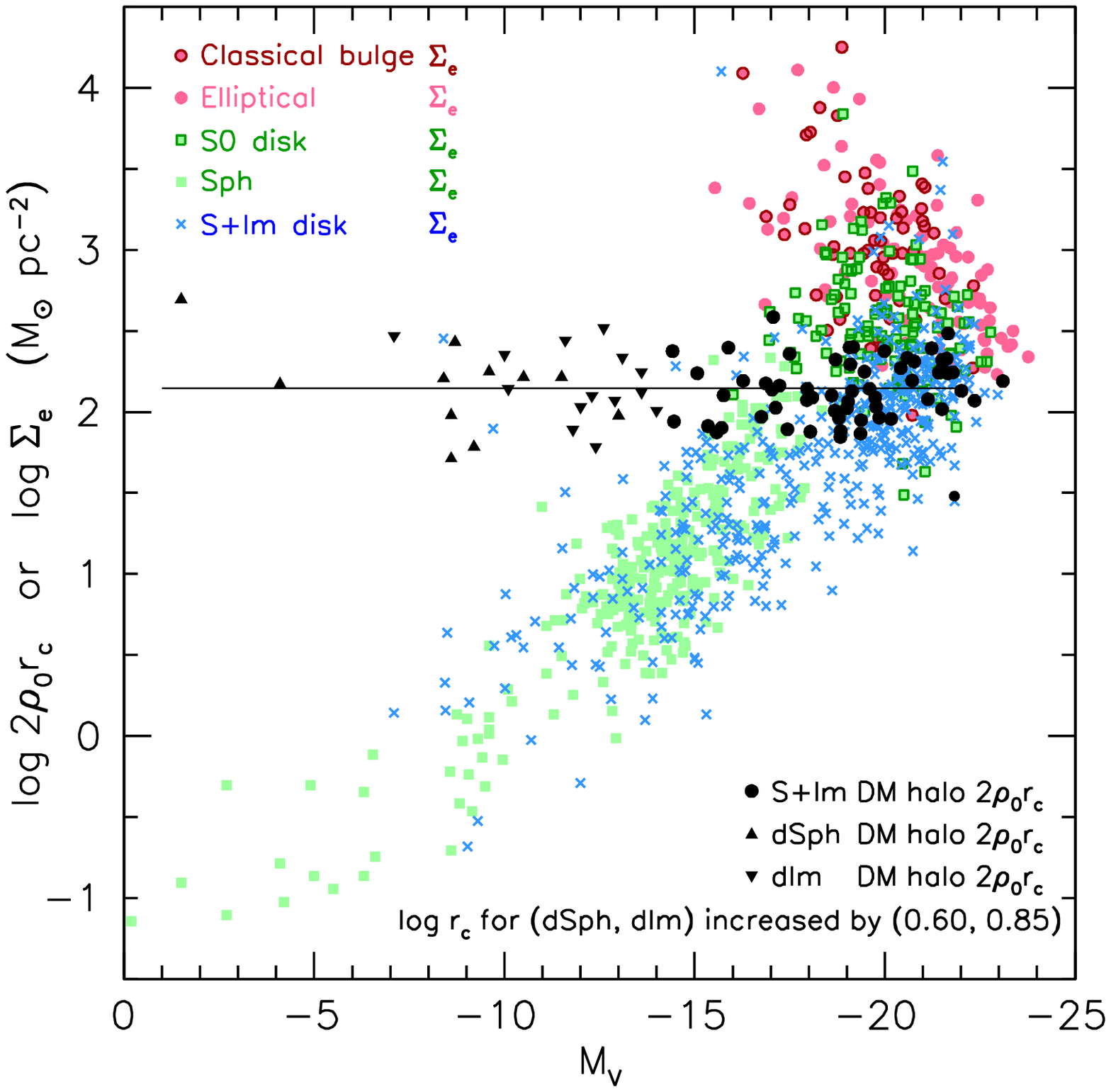}

\noindent {\bf Fig.~6}~~Comparison of dark matter (DM) halo parameters from Kormendy \& Freeman (2014) with visible matter 
parameters from Kormendy \& Bender (2012).  
DM parameters are from maximum-disk rotation curve decompositions ({\it black circles\/}) or from cored isothermal halo 
models applied to the dispersion profiles of dSph galaxies ({\it black triangles\/}) or to the $V \propto r$ rotation curves
and velocity dispersions of H{\ts}I in dIm galaxies ({\it upside-down black triangles\/}) (see Kormendy \& Freeman 2014, the
source of this figure).  Central projected densities are plotted for DM halos; effective surface densities $\Sigma_e = \Sigma(r_e)$
are shown for visible components.  Here $r_e$ is the radius that contains half of the light of the component.
Surface brightnesses are converted to stellar surface densities  using mass-to-light ratios 
$M/L_V = 8$ for ellipticals,
         5  for classical bulges and S0 disks, and
         2  for spiral galaxy disks, Im galaxies, and Sph galaxies.

\eject

\noindent together form a sequence of increasing dissipation during the formation of smaller galaxies.   
For $M_V < -18$ galaxies of all kinds, effective densities in stars are similar to DM densities at and 
interior to the same radius.  For Sc--Im systems, this is by construction a consequence (1) of using maximum-disk
decompositions and (2) of the ``rotation curve conspiracy'' (van Albada \& Sancisi 1986), i.{\ts}e., the observation
that rotation curves of giant galaxies are roughly flat and featureless, so the parts of galaxies
that are controlled by dark matter are not easily distinguished from the parts that are controlled by visible matter 
or even the parts that are controlled by different components in the visible matter (Figure 4).  Caveat:
for bulges and ellipticals, high baryon densities at $r \ll r_e$ may pull on DM halos enough to
increase their central densities over the values for Sc{\ts}--{\ts}Im galaxies that are shown in Figure 6.
But bulges and ellipticals have {\it central\/} projected densities that are more than 3 dex higher than
the effective densities shown in Figure 6.  So the central parts of early-type galaxies are {\it very\/}
baryon-dominated.  Even the central densities of disks are 0.7 dex (for an exponential) higher
than the effective densities shown in Fig.~6.  So even pure disks are moderately dominated by visible
matter near their centers.  Both results are qualitatively as expected: Visible matter needs to dissipate,
sink inside the DM, and become self-gravitating enough to form stars and visible galaxies.  And
a great deal of dissipation happens in the wet mergers that make normal ellipticals (KFCB):~their densities 
rise above DM densities by larger amounts at fainter $M_V$.

      The important point here is this: At $M_V > -18$, tinier dwarfs are more DM dominated, until by 
$M_V$\ts\gapprox\ts$-10$, they are essentially dark galaxies with just enough of a frosting of stars so
that they can be detected.  I emphasize two important 
points: (1) The differences between dIm and dSph galaxies in all parameter correlations shown in this paper are
small.  Whether or not a galaxy retains cold gas and can still form stars in today's Universe is a second-order
effect.  This argues -- as Dekel \& Silk (1986) already emphasized -- that the primary effect that engineers the
parameter correlations is supernova-driven baryon blowout or another process (such as a failure to capture baryons 
before cosmic reionization) that has the same effect.  (2) Kormendy \& Freeman (2014) suggest that there exists a 
large population of tiny halos that
are essentially completely dark and that the discoverable galaxies at $M_V$\ts\gapprox\ts$-13$ represent a smaller and
smaller fraction of tinier DM halos.  This has been suggested as the solution to the problem
that the fluctuation spectrum of cold dark matter predicts more dwarfs than are observed in environments
like the Local Group (e.{\ts}g.,
Moore et al.~1999; 
Klypin et al.~1999).

\vfill

\section{Conclusions}
\label{sec:5}

      Kormendy \& Bender (2012) propose a parallel sequence galaxy classification (Figure~1~here) in which Sph galaxies 
appear as bulgeless S0s juxtaposed with Im galaxies.  In reality, their progenitors can include late-type spirals, but
the Fig.\ts1 tuning fork is designed for simplicity.  S0$+$Sph galaxies are suggested to be 
star-formation-quenched descendants of S$+$Im galaxies.  It seems essentially guaranteed that all  
transformation processes discussed in \S3 are important.  Moreover, although there is good agreement between 
structural parameter correlations for (1) present-day Sphs$+$S0 disks and (2) present-day Im$+$S disks, it is of course 
not guaranteed that the progenitors of Sph and S0 galaxies were exactly like present-day late-type galaxies.  The latter are, 
after all, survivors.  The reasons can partly involve stochastic evolution, but they may also involve differences in internal 
structure and/or environment.

      These ideas are slowly gaining acceptance.  An observationally biased but still
incomplete list of papers includes
Grebel{\ts}et{\ts}al.\ts(2003);
van Zee{\ts}et{\ts}al.\ts(2004);
Mayer{\ts}et{\ts}al.\ts(2006); 
Boselli{\ts}et al.\ts(2008a,{\ts}b);
Tolstoy{\ts}et{\ts}al.\ts(2009);
\hbox{Ry\'s{\ts}et{\ts}al.\ts(2013, 2014); and
Janz \etal (2013).}
KFCB and Kormendy\ts\&{\ts}Bender\ts(2012) address remaining controversies. 

\eject

      From the perspective of this conference, the important -- and robust! -- conclusion is this: \vskip 4pt

      {\it Dwarf spiral, irregular, and spheroidal galaxies are not new and special kinds of galaxies}, as 
has often been suggested.  Rather: {\it There is a complete continuity in structural parameter
scaling relations} (Figures 2\ts--\ts4 and 6), {\it kinematics and dynamics, star formation properties, and
metallicity distributions between the disks of giant galaxies and all three subsets of dwarf galaxies.  
Fainter than $M_V \sim -18$, the properties of galaxies change as their gravitational potential wells get more shallow;
this is qualitatively consistent with the increased importance at lower masses of internal and environmental 
transformation processes.  Also, many aspects of galaxy evolution get more stochastic in smaller galaxies.
For example, star formation gets more bursty.  Many quantitative details remain to be worked out.  But this is
engineering.  A~big-picture view of the evolution of dwarf galaxies seems comfortably in place:} \vskip 4pt

      {\it All forms of violence -- including supernova energy feedback, dynamical harassment, and external ram-pressure
stripping -- get more important for smaller galaxies.  The main result is that smaller galaxies lose more
of their baryons or never acquire them.  Smaller dwarfs are more dominated by dark matter.  Whether
or not they retain enough cold gas to feed some star formation is very much a second-order effect. 
Baryon loss at low halo masses may be so severe that we discover only a small fraction of the smallest, 
essentially dark halos.  This can reconcile the observed scarcity of dwarf galaxies in field environments 
with our expectations of large numbers of dwarfs that are predicted by the cold dark matter fluctuation spectrum}.

\section{Acknowledgments}

      It is a pleasure to thank Ralf Bender and Ken Freeman for their hospitality and support during my visits
and for always enjoyable collaborations.~My attendance at the Seychelles conference was supported by the 
Max-Planck-Institute for Extraterrestrial Physics and by the Observatory of the Ludwig-Maximilians-University, Munich.  
This paper was supported by the Curtis T. Vaughan, Jr. Centennial Chair in Astronomy at the University of Texas.

\section{References}

\parindent=0pt

Bedregal, A. G., Arag\'on-Salamanca, A., \& Merrifield, M. R. 2006, MNRAS, 373, 1125


B\"ohringer, H., \etal 1994, Nature, 368, 828  

Boselli, A., Boissier, S., Cortese, L., \& Gavazzi, G. 2008a, ApJ, 674, 742 

Boselli, A., Boissier, S., Cortese, L., \& Gavazzi, G. 2008b, A\&A, 489, 1015

Cappellari, M., \etal 2011, MNRAS, 416, 1680

Cayette, V., Kotanyi, C., Balkowski, C., \& van Gorkom, J.~H.~1994, AJ, 107, 1003

Cayette, V., van Gorkom, J.~H., Balkowski, C., \& Kotanyi, C.~1990, AJ, 100, 604

Chung, A., van Gorkom, J.~H., Kenney, J.~D.~P., Crowl, H., \& Vollmer, B.~2009, AJ,\ts138,\ts1741  

Chung, A., van Gorkom, J.~H., Kenney, J.~D.~P., \& Vollmer, B.~2007, ApJ, 659, L115

Courteau, S., \etal 2007, ApJ, 671, 203

Dalcanton, J. J., Yoachim, P., \& Bernstein, R. A. 2004, ApJ, 608, 189

Dav\'e, R., \etal 2001, ApJ, 552, 473

Dekel, A., \& Silk, J. 1986, ApJ, 303, 39

Einasto, J., Saar, E., Kaasik, A., \& Chernin, A. D. 1974, Nature, 252, 111

Faber, S.~M., \& Lin, D.~N.~C.~1983, ApJ, 266, L17

Freeman, K.~C.~1970, ApJ, 160, 811

Gatto, A., \etal 2013, MNRAS, 433, 2749

Grebel, E. K., Gallagher, J. S., \& Harbeck, D. 2003, AJ, 125, 1926

Gunn, J.~E., \& Gott, J.~R.~1972, ApJ, 176, 1


Gursky, H., \etal 1971, ApJ, 167, L81

Janz, J., \etal 2013, arXiv:1308.6496  

Kenney, J.~D.~P., Tal, T., Crowl, H.~H., Feldmeier, J., \& Jacoby, G.~H.~2008, ApJ, 687, L69

Kenney, J.~D.~P., van Gorkom, J.~H., \& Vollmer, B.~2004, AJ, 127, 3361  

Klypin, A., Kravtsov, A.~V., Valenzuela, O., \& Prada, F.~1999, ApJ, 522, 82

Kormendy, J.~1985, ApJ, 295, 73

Kormendy, J.~1987, in Nearly Normal Galaxies: From the Planck Time
             to the Present, ed.~S.~M.~Faber \phantom{~}\quad (New York: Springer), 163

Kormendy, J., \& Bender, R. 2012, ApJS, 198, 2~~(They list references in the keys of Figures\ts2{\ts}and\ts3.)

Kormendy, J., Fisher, D.~B., Cornell, M.~E., \& Bender, R.~2009, ApJS, 182, 216 (KFCB)

Kormendy, J., \& Freeman, K. C. 2014, ApJ, in preparation

Krajnovi\'c, D., \etal 2011, MNRAS, 414, 2923

Lake, G., Katz, N., \& Moore, B. 1998, ApJ, 495, 152

Larson, R. B. 1974, MNRAS, 169, 229

Larson, R. B., Tinsley, B. M., \& Caldwell, C. N. 1980, ApJ, 237, 692

Laurikainen E., Salo H., Buta R., Knapen J. H., \& Comer\'on S., 2010, MNRAS, 405, 1089

Lin, D.~N.~C., \& Faber, S.~M.~1983, ApJ, 266, L21

Mateo, M. 1998, ARA\&A, 36, 435

Mayer, L., Mastropietro, C., Wadsley, J., Stadel, J., \& Moore, B. 2006, MNRAS, 369, 1021

Mayer, L., \etal 2001a, ApJ, 547, L123

Mayer, L., \etal 2001b, ApJ, 559, 754

Meekins, J.~F., Fritz, G., Chubb, T.~A., Friedman, H., \& Henry, R.~C.~1971, Nature, 231, 107

Moore, B., Katz, N., Lake, G., Dressler, A., \& Oemler, A. 1996, Nature, 379, 613

Moore, B., Lake, G., \& Katz, N. 1998, ApJ, 495, 139

Moore, B., \etal 1999, ApJ, 524, L19

Oosterloo, T., \& van Gorkom, J.~2005, A\&A, 437, L19  

Pavel, J., Combes, F., Cortese, L., Sun, M., \& Kenney, J. D. P. 2014, arXiv:1403.2328

Ry\'s, A., Falc\'on-Barroso, J., \& van de Ven, G. 2013, MNRAS, 428, 2980

Ry\'s, A., van de Ven, G., \& Falc\'on-Barroso, J. 2014, MNRAS, 439, 284

Saito, M. 1979, PASJ, 31, 193

Sandage, A., \& Binggeli, B. 1984, AJ, 89, 919

Sivanandam, S., Rieke, M. J., \& Rieke, G. H. 2010, ApJ, 717, 147

Sun, M., Donahue, M., \& Voit, G. M. 2007, ApJ, 671, 190

Sun, M., \etal 2006, ApJ, 637, L81

Sun, M., \etal 2010, ApJ, 708, 946

Tolstoy, E., Hill, V., \& Tosi, M. 2009, ARA\&A, 47, 371

Tully, R. B., \& Fisher, J. R. 1977, A\&A, 54, 661

van Albada, T.~S. \& Sancisi, R.~1986, Phil.~Trans.~R.~Soc.~London A, 320, 447

van den Bergh, S. 1976, ApJ, 206, 883  

van den Bergh, S.~1994a, AJ, 107, 1328  

van den Bergh, S.~1994b, ApJ, 428, 617  

van Zee, L., Skillman, E. D., \& Haynes, M. P. 2004, AJ, 128, 121

Woudt, P.{\ts}A., Kraan-Korteweg, R.{\ts}C., Lucey, J., Fairall, A.{\ts}P., \& Moore, S.{\ts}A.{\ts}W. 2008, MNRAS, \phantom{~}\quad 383, 445
 
\vfill\eject

%
%


\end{document}